\documentclass{PoS}
\pdfoutput=1 

\usepackage{color}
\usepackage{amsmath}
\usepackage{graphicx}

\graphicspath{ {figs/} }

\newcommand{\lsim}{\;\rlap{\lower 3.5 pt \hbox{$\mathchar \sim$}} \raise 1pt
 \hbox {$<$}\;}

\title{
\vspace*{-5em}
\mbox{}\hfill \mbox{\small\sc TTP19-042, P3H-19-048}\\
\vspace*{5em}
NNLO real corrections to $gg\to HH$ in the large-$m_t$ limit}

\ShortTitle{NNLO real corrections to $gg\to HH$ in the large-$m_t$ limit}

\author{
  Joshua Davies\\
  Karlsruhe Institute of Technology (KIT), Karlsruhe, Germany\\
  E-mail: \email{joshua.davies@kit.edu}}
\author{
  Florian Herren\\
  Karlsruhe Institute of Technology (KIT), Karlsruhe, Germany\\
  E-mail: \email{florian.herren@kit.edu}}
\author{
  Go Mishima\\
  Karlsruhe Institute of Technology (KIT), Karlsruhe, Germany\\
  E-mail: \email{go.mishima@kit.edu}}
\author{
  \speaker{Matthias Steinhauser}
  \\
  Karlsruhe Institute of Technology (KIT), Karlsruhe, Germany\\
  E-mail: \email{matthias.steinhauser@kit.edu}}

\abstract{In this contribution we consider NNLO real radiation
  corrections to the total cross section for Higgs boson pair
  production in gluon fusion. Special emphasis is put
  on the cross check of the asymptotic expansion in the inverse top
  quark mass.
}

\FullConference{14th International Symposium on Radiative Corrections (RADCOR2019)\\ 
  9-13 September 2019\\
  Palais des Papes, Avignon, France}

\begin{document}


\section{\label{sec::intro}Introduction}

After the discovery of the Higgs boson~\cite{Aad:2012tfa,Chatrchyan:2012xdj} all parameters and
couplings of the Standard Model (SM) are fixed. In particular, the
coupling strength for the interaction of three and four Higgs bosons is given
by $\lambda = m_H^2/(2v^2) \approx 0.13$ where $m_H$ is the Higgs boson
mass and $v$ is the vacuum expectation value. However, many beyond-the-SM theories implement a different
scalar sector. It is thus desirable to obtain independent information
about the Higgs boson self coupling from experimental measurements.  A
promising process in this context is double Higgs boson production.

In the recent years many higher order corrections to
Higgs boson pair production have become available, in particular
for the numerically most important channel $gg\to HH$.
In this proceedings contribution we refrain from
providing a detailed listing of all relevant works
but refer to a recent review~\cite{DiMicco:2019ngk}
and references cited therein.

The aim of this contribution is to summarize the
work~\cite{Davies:2019xzc} and provide further technical details on
the extension to the contribution with two closed top quark loops.
In particular, we confront the ``building-block-approach''
with the naive use of asymptotic expansion.


\section{Setup}

\begin{figure}[b]
  \begin{center}
    \begin{tabular}{cccc}
      \includegraphics[width=0.22\textwidth]{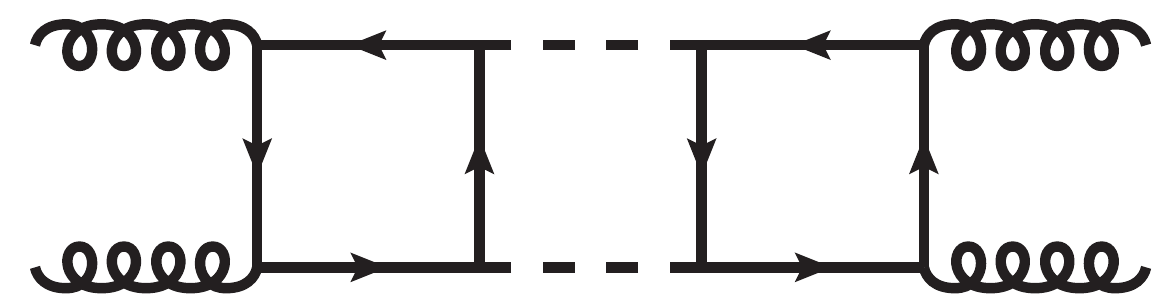}     &
      \includegraphics[width=0.22\textwidth]{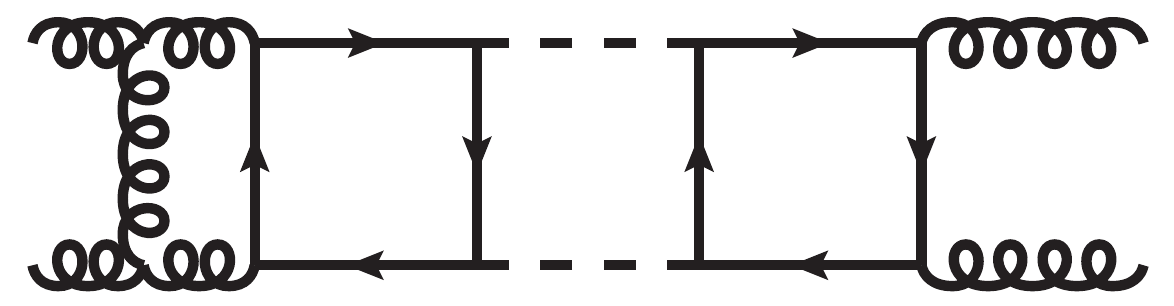} &
      \raisebox{-1em}{\includegraphics[width=0.22\textwidth]{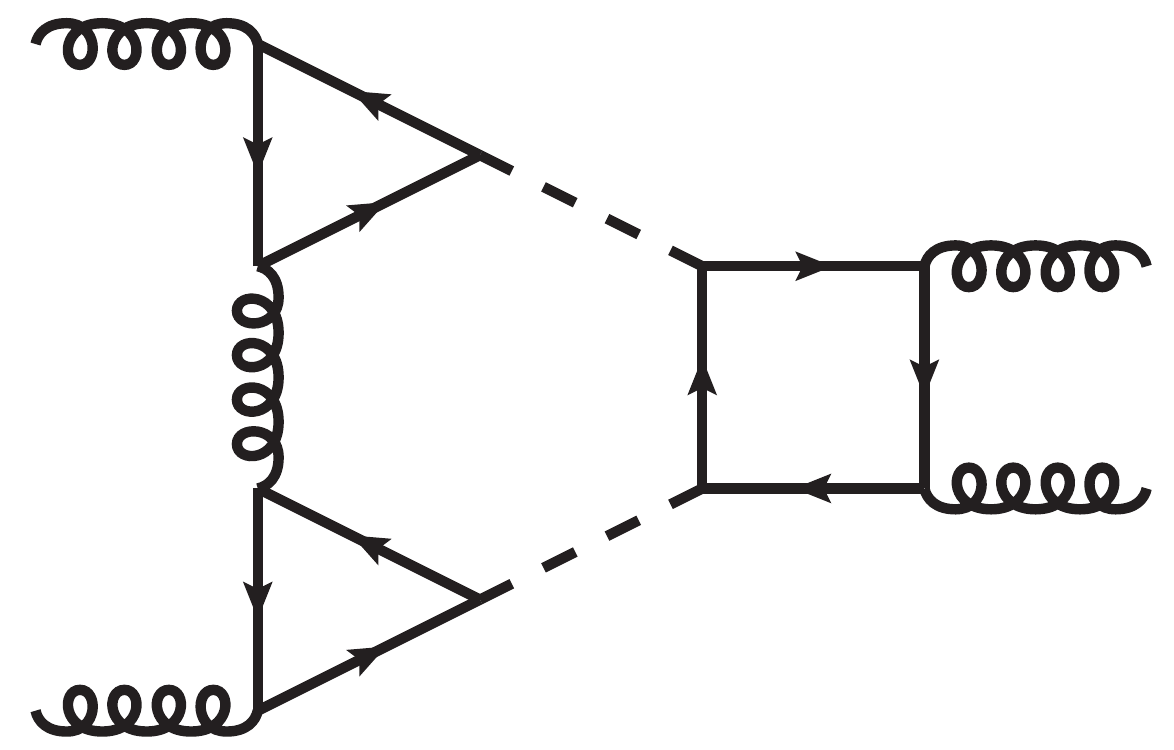}} &
      \includegraphics[width=0.22\textwidth]{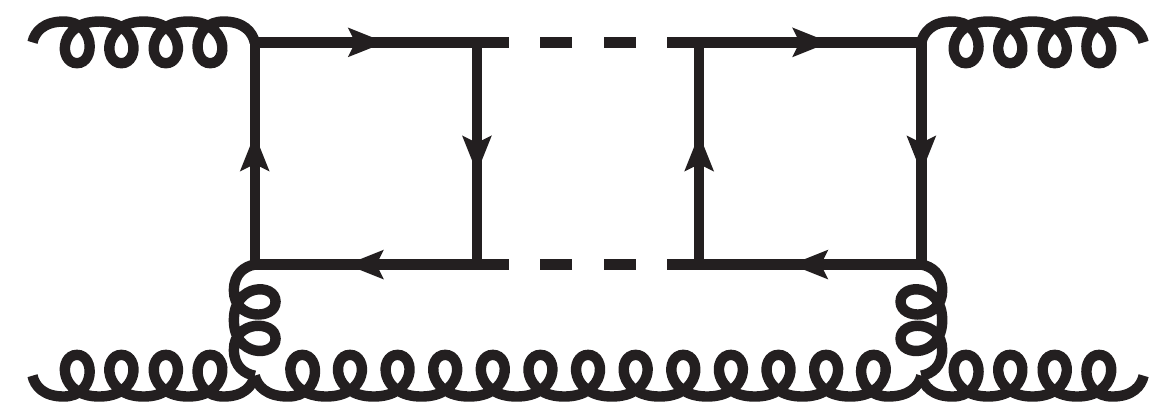} \\
    \end{tabular}
  \end{center}
  \caption{\label{fig::LO_NLO} One LO and three NLO sample Feynman diagrams for
    $gg\to gg$. Solid, dashed and curly lines represent quarks, Higgs
    bosons and gluons, respectively.  The contributions to the Higgs
    boson pair production cross section are obtained by considering
    cuts which involve at least two Higgs bosons.
    The third diagram represents the $n_h^3$ contribution
    at NLO.}
\end{figure}

\begin{figure}[t]
  \begin{center}
    \begin{tabular}{ccc}
      \raisebox{-1.8em}{\includegraphics[width=0.22\textwidth]{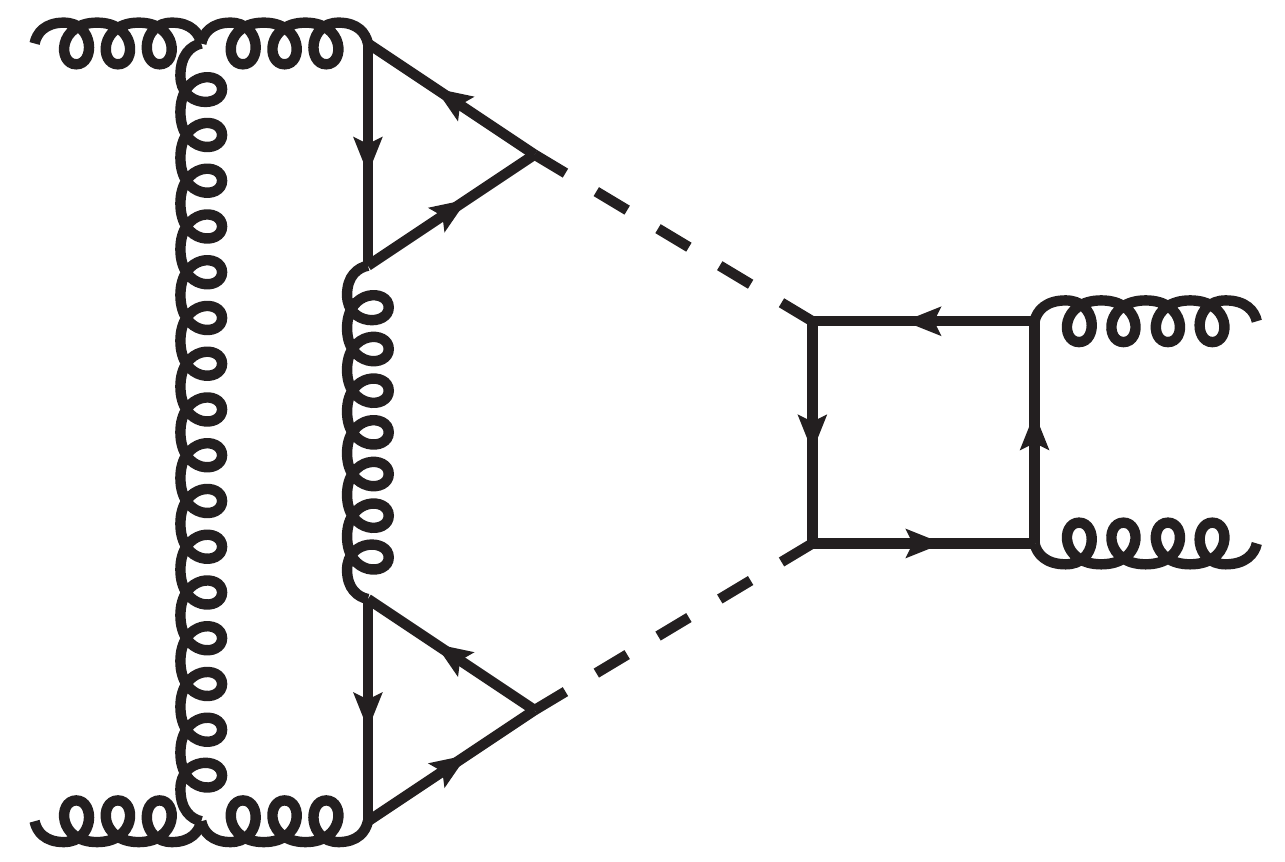}}
      &
      \includegraphics[width=0.22\textwidth]{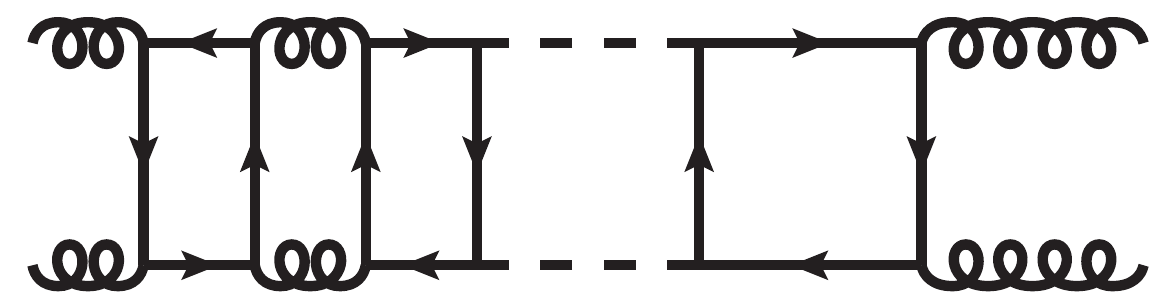} 
      &
      \raisebox{-1.3em}{\includegraphics[width=0.22\textwidth]{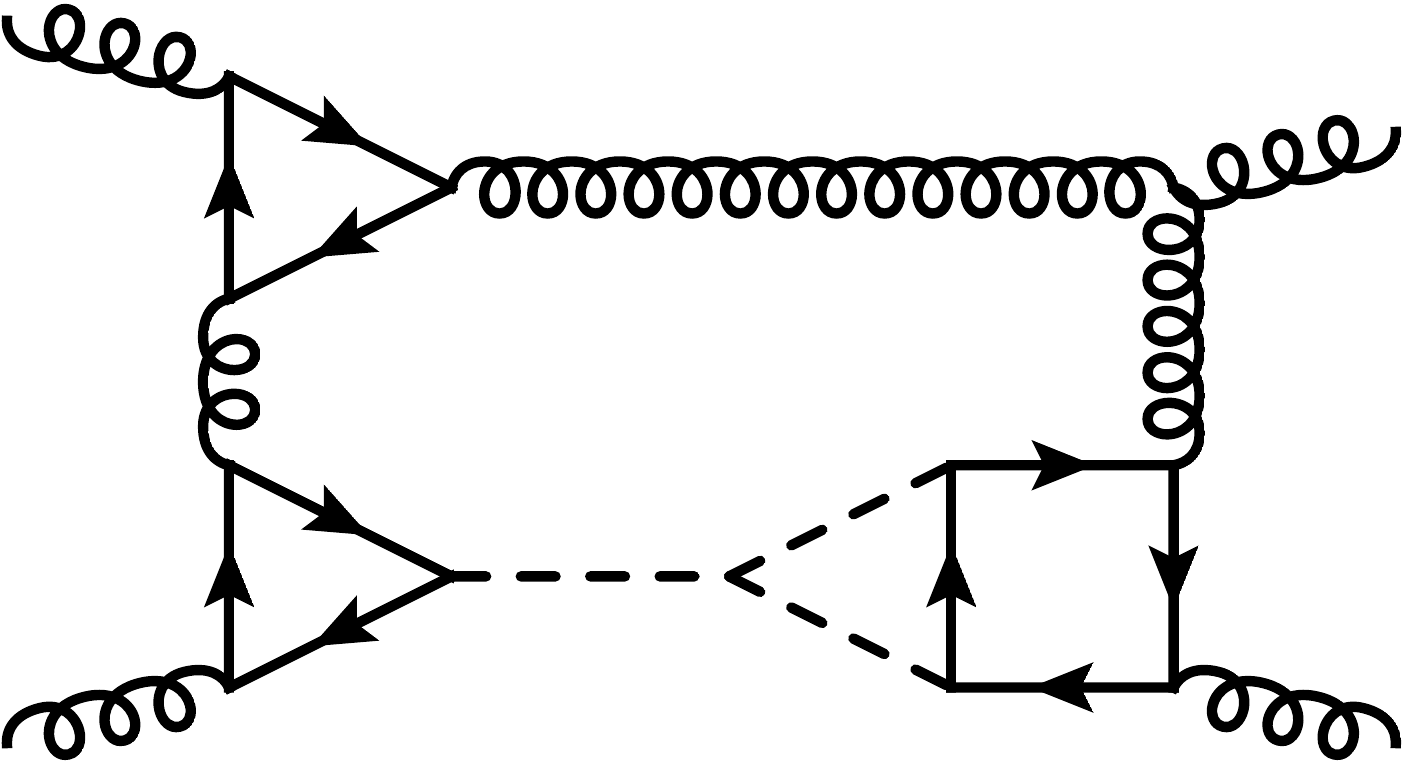}}
      \\
      \raisebox{-1em}{\includegraphics[width=0.22\textwidth]{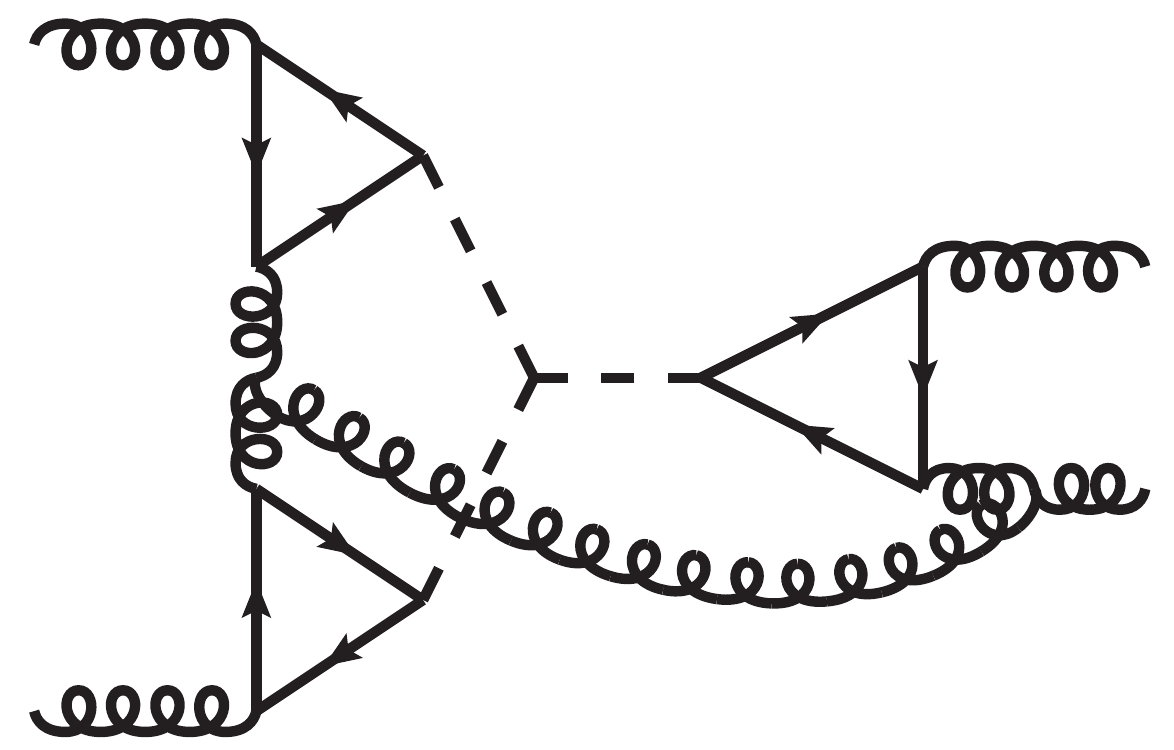}}
      &
      \includegraphics[width=0.22\textwidth]{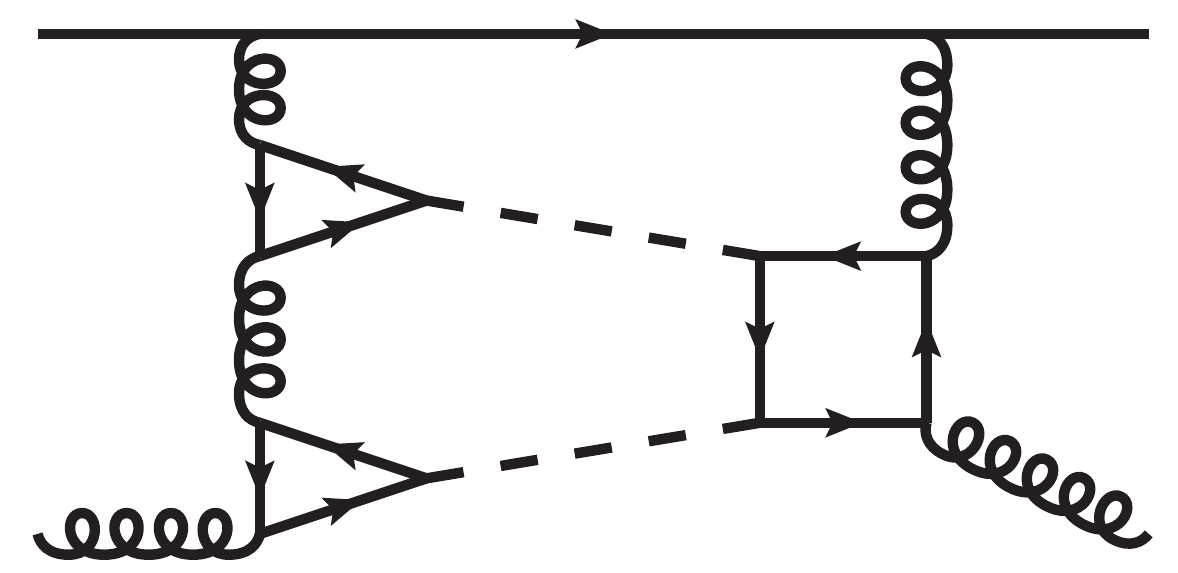}
      &
      \raisebox{1em}{\includegraphics[width=0.22\textwidth]{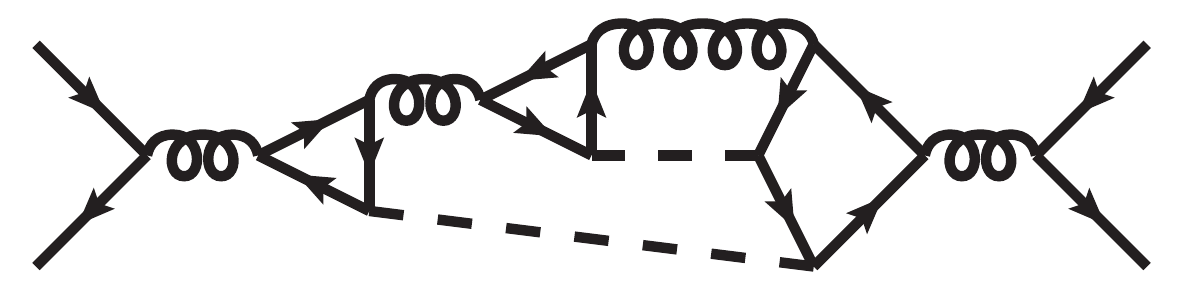}}
    \end{tabular}
  \end{center}
  \caption{\label{fig::sample_FDs} Sample NNLO
    Feynman diagrams for $ij\to
    ij$ with $i,j\in\{g,q\}$. Solid, dashed and curly lines represent
    quarks, Higgs bosons and gluons, respectively.
    The contributions to the Higgs boson
    pair production cross section are obtained by considering cuts
    which involve at least two Higgs bosons.
    All diagrams are  $n_h^3$ contributions except
    the second and the third diagram in the first row, 
    which are not included in our final result. They contain a closed top
    quark loop without a coupling to the Higgs boson.}
\end{figure}

In the following we briefly describe the individual
steps which we follow in order to arrive at analytic
results for the partonic cross section for $gg\to HH$.
We are interested in computing the imaginary part of the
forward-scattering amplitude\footnote{For simplicity we discuss here
  the amplitude $gg\to gg$. At NLO there are in addition the $qg$ and
    $q\bar{q}$ channels and
    and NNLO also the $qq$ and $qq^\prime$ channels.}
$g(q_1)g(q_2)\to g(q_1)g(q_2)$ due to
cuts involving two Higgs bosons and possibly further light
(anti-)quarks or gluons.  Note that at LO, NLO and NNLO this leads
to three-, four- and five-loop diagrams; all of them have closed top
quark loops both left and right of the cut.  In fact, we can classify
the diagrams according to the number of closed top quark loops which
involve at least one coupling to a Higgs boson. 
The real radiation corrections at NNLO have
either two or three, which we denote by $n_h^2$ and $n_h^3$
contributions, respectively. Results for the $n_h^3$ contribution have
been published in Ref.~\cite{Davies:2019xzc}.
At LO
there are only $n_h^2$ diagrams. $n_h^3$ contributions
appear for the first time at NLO as virtual corrections
as can be seen in Fig.~\ref{fig::LO_NLO}
(NNLO $n_h^3$ diagrams are shown in Fig.~\ref{fig::sample_FDs}.)
The virtual corrections at NNLO also have $n_h^4$ terms.

We generate the amplitudes of the individual Feynman diagrams using
{\tt qgraf}~\cite{Nogueira:1991ex}.  After specifying the particle
content this leads to $16.6\times 10^6$ diagrams. However, many
of them have no relevant cuts.  For example, in many cases the Higgs
bosons are generated in the $t$ instead of the $s$ channel.  We thus
apply additional scripts to select the contributions containing the
relevant cuts which significantly reduces the number of diagrams
to $0.16\times 10^6$;
$12,114$ of them contribute to the $n_h^3$ terms.

Instead of generating five-loop amplitudes (at NNLO) it is possible to
interpret the subdiagrams containing the top quark as effective
vertices\footnote{Note that these effective vertices are generated at
  the integrand level and that there is no effective Lagrange
  density. For our expansion depth, the construction of this would
  require operators up to dimension twelve.} which mediate the
coupling of up to two Higgs bosons, up to four gluons and up to one
quark-anti-quark pair.  One can pre-compute the large-$m_t$ expansion
of these one- and two-loop tadpole integrals using using {\tt
  MATAD}~\cite{Steinhauser:2000ry} and store the results to disk.
Afterwards we generate $g(q_1)g(q_2)\to g(q_1)g(q_2)$ amplitudes up to
three loops using effective vertices in the {\tt qgraf} input file. In
the course of the calculation the effective vertices are replaced by
the pre-computed results for the top quark loops.  We call this
approach the ``building-block approach'' and provide more details in
Section~\ref{sec::ae}.

The modified {\tt qgraf} output is then processed by {\tt q2e} and
{\tt exp}~\cite{Harlander:1997zb,Seidensticker:1999bb,q2eexp}, which
generate {\tt FORM}~\cite{Ruijl:2017dtg} code for the amplitudes and
map them onto the predefined integral families. We compute the colour
factors of the diagrams using {\tt color}~\cite{vanRitbergen:1998pn}.

Next, we must apply partial fraction decompositions to
arrive at a unique set of integral families.  This is necessary since
we consider four-point functions but have only two independent
external momenta. In fact, for our kinematic configuration we have
3, 7 and 12 independent kinematic invariants (and thus
scalar functions with 3, 7 and 12 indices) at one, two and three
loops, respectively.
A subsequent reduction, which we perform
with the help of {\tt LiteRed}~\cite{Lee:2012cn,Lee:2013mka},
leads to master integrals which depend on
\begin{eqnarray}
  x &=& \frac{m_H^2}{s} \,.
\end{eqnarray}
In the physical region we have $0 < x < 1/4$.
In the next section we discuss the evaluation of the master integrals
for the NNLO $n_h^3$ term.


\section{$n_h^3$ contribution}

\begin{figure}[t]
  \begin{center}
    \includegraphics[width=.5\textwidth]{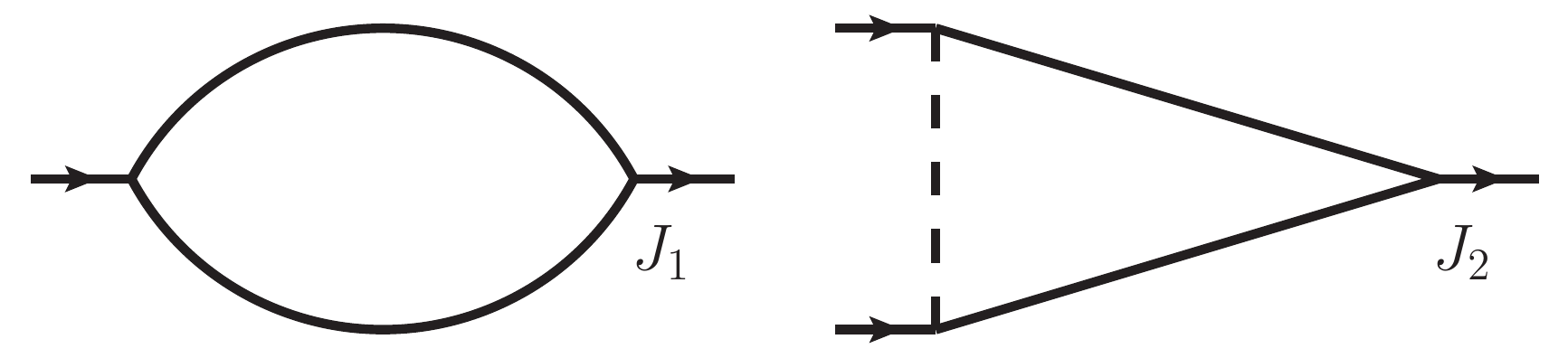}
    \\
    \includegraphics[width=\textwidth]{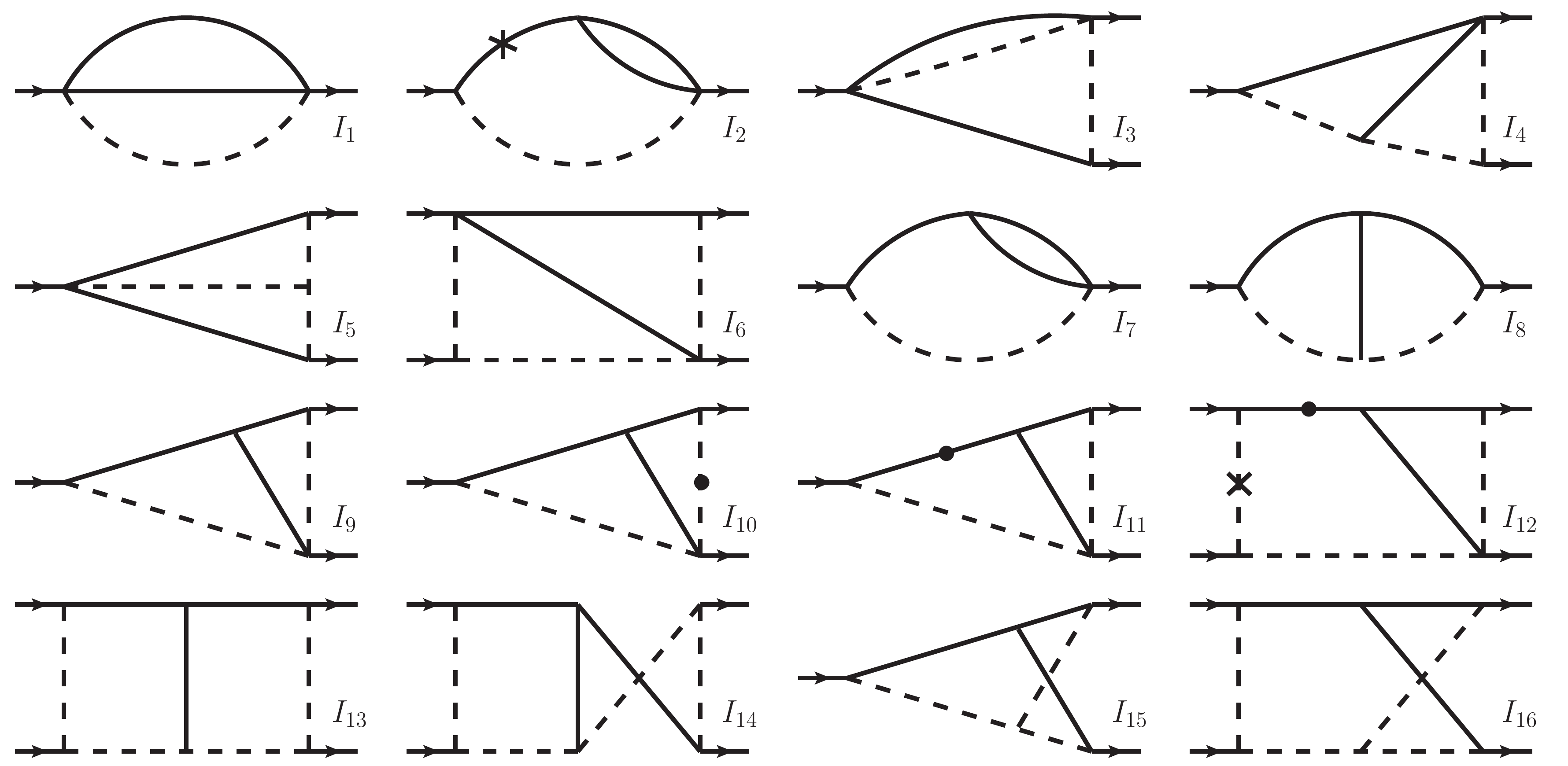}
  \end{center}
  \caption{\label{fig::masters}
    One- (first row) and two-loop
    master integrals.  Solid and dashed lines represent massive and
    massless propagators. Dots and crosses denote squared and
    inverse propagators, respectively. It is understood that the momenta $q_1$ and
    $q_2$ enter the diagrams on the left and leave them on the right
    in the upper and lower lines, respectively. For the external momenta we
    have $q_1^2=q_2^2=0$ and $(q_1+q_2)^2=s$.}
\end{figure}

Sample Feynman diagrams which have to be considered for the NNLO $n_h^3$
contribution are shown in Fig.~\ref{fig::sample_FDs}.  The virtual
corrections have been computed in
Refs.~\cite{Grigo:2015dia,Davies:2019djw}. As for the real
corrections, each diagram in the second
row is a representative of one of the three partonic channels, $gg$,
$qg$ and $q\bar{q}$.  After applying the steps described in the
previous section we can express the $gg\to gg$ amplitude as a linear
combination of 2~one-loop and 16~two-loop master integrals which we
show in Fig.~\ref{fig::masters}.  They depend on $x$ and analytic
results are obtained using the method of differential
equations~\cite{Kotikov:1990kg,Kotikov:1991hm,Kotikov:1991pm}.

The master integrals shown in Fig.~\ref{fig::masters} can be
transformed into the so-called $\epsilon$ form of the
differential equation, which has the particularly simple
form\footnote{See Eq.~(\ref{eq::y}) for the relation between $x$ and $y$.}
\begin{eqnarray}
  \frac{{\rm d}I}{{\rm d}y} &=& \epsilon \sum_i \frac{A_i}{y-y_i} I 
  \,,
  \label{eq::ep_form}
\end{eqnarray}
where $I$ is the vector of master integrals and $A_i$ are square
matrices with constant (i.e., independent of $y$ and $\epsilon$)
matrix elements.  Due to the fact that in Eq.~(\ref{eq::ep_form})
$\epsilon$ factorizes and the $y$ dependence only occurs in form of simple poles
it is straightforward to construct solutions of $I$ in terms of
iterated integrals (Goncharov polylogarithms~\cite{Goncharov:1998kja})
provided we have boundary conditions for $I$ for some value of
$y$.  We have chosen the so-called soft limit which corresponds to
\begin{eqnarray}
  \delta\equiv \sqrt{1-4x}\to0\,,
  \quad \mbox{i.e.} \quad x\to 1/4 \quad \mbox{and} \quad y\to -1.
\end{eqnarray}
In this contribution we refrain from
discussing details about the computation of the boundary values;
they can be found in Ref.~\cite{Davies:2019xzc}.

In order to arrive at Eq.~(\ref{eq::ep_form}) one has to
apply ideas developed in Refs.~\cite{Henn:2013pwa,Lee:2014ioa}
which help to transform the original system of differential equations
into Eq.~(\ref{eq::ep_form}). In practice, we use
the program {\tt Epsilon}~\cite{Prausa:2017ltv} which is based
on the algorithm provided in Ref.~\cite{Lee:2014ioa}.
In an intermediate step we observe poles in $x$ at the positions
$x = \{0,1/4,1,r_1=\exp(i\pi/3),r_2=\exp(-i\pi/3),-1/3\}$
A closer inspection of the corresponding matrices
suggests the variable transformation
\begin{eqnarray}
  y &=& \frac{\sqrt{1-4x}-1}{\sqrt{1-4x}+1},~-1 < y < 0\,,
  \label{eq::y}
\end{eqnarray}
which leads to an $\epsilon$ form for the first 14 (out of 16) master integrals,
as discussed in Ref.~\cite{Davies:2019xzc}. The explicit construction
of an $\epsilon$ form for the remaining two master integrals
can be avoided since at most their leading $\epsilon$ term enters the
physical result.

As an alternative to the exact solution of Eq.~(\ref{eq::ep_form}) in
terms of generalized logarithms one can use the differential equation
together with the boundary conditions to construct many terms
of an expansion in $\delta$. We managed to obtain, without much
difficulty, more than 500 expansion terms; these are more than sufficient
for all practical purposes.

\begin{figure}[t]
  \begin{center}
    \includegraphics[viewport=50 150 550 730,width=.9\textwidth]{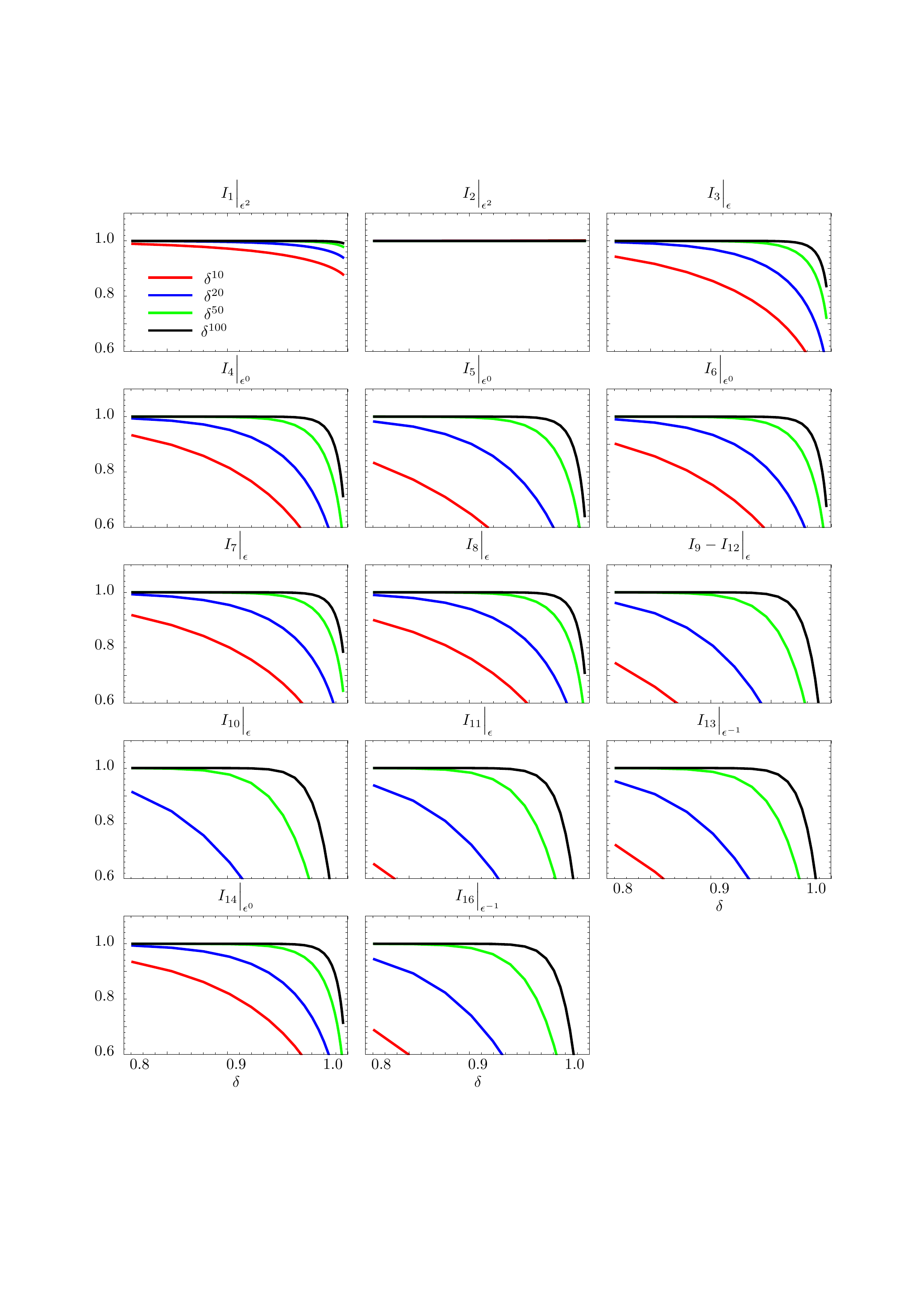}
  \end{center}
  \caption{\label{fig::mi_ep0}
    Two-loop master integrals of
    Fig.~\ref{fig::masters} as a function of $\delta$.
    We show the highest $\epsilon$ term which enters the finite result
    of the physical amplitude. The curves include expansions in $\delta$  up
    to order $\delta^{10}$, $\delta^{20}$, $\delta^{50}$ and $\delta^{100}$
    normalized to the exact result. Note that $I_{15}$ is finite
    and enters the amplitude with a prefactor proportional to $\epsilon$.
    Note that in the case
    of $I_2$ the higher order $\delta$ terms rapidly become quite small
    which is the reason for the good convergence.}
\end{figure}

In Fig.~\ref{fig::mi_ep0} we show, for each master
integral, the highest $\epsilon$ term needed for the physical amplitude.
We normalize the different expansions to the exact result
and plot the ratio as a function of $\delta$.
Note that $I_{15}$ does not contribute to the amplitude. However,
it is needed for the computation of $I_{16}$ since $I_{15}$ is 
present in one of its subsectors.
In all cases, one observes good agreement between the exact result and the
expansion (including terms to $\delta^{50}$) in the region $\delta\le0.9$.
We note that $\delta=0.9$ corresponds to $\sqrt{s}\approx 800$~GeV.
This has to be compared with the validity range of the large-$m_t$
expansion (see below) which is for $\sqrt{s}\lsim 2m_t \approx 350$~GeV.

\begin{figure}[t]
    \begin{center}
        \includegraphics[width=0.95\textwidth]{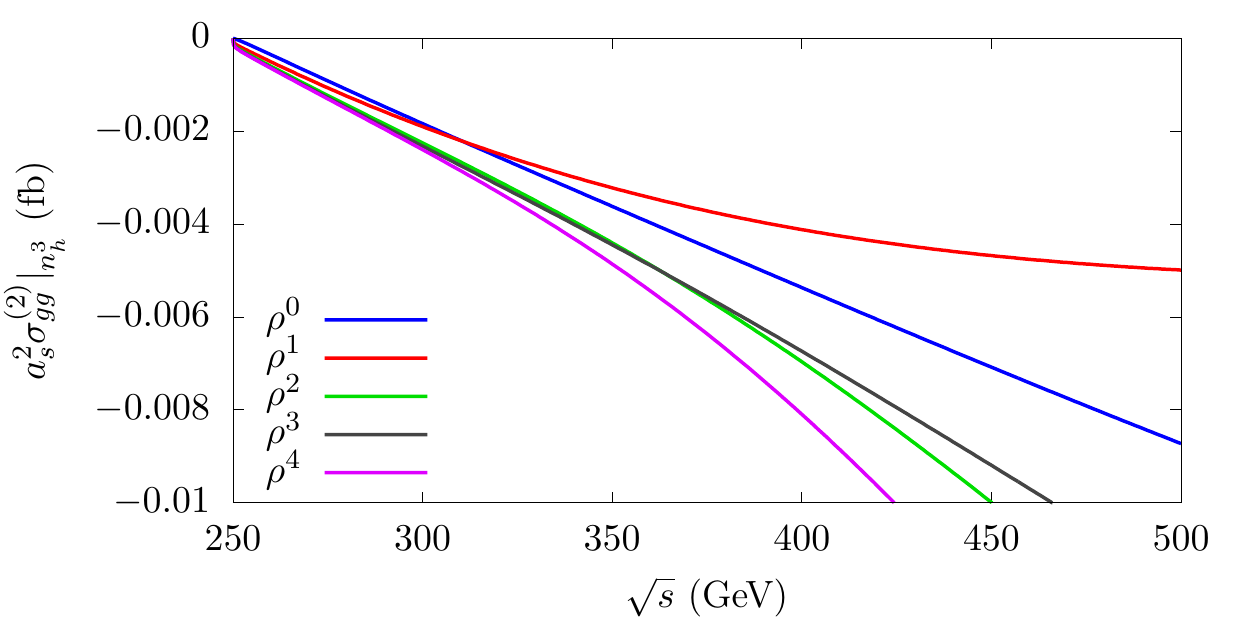}
        \caption{\label{fig::gg}Partonic NNLO $n_h^3$ cross sections
          as a function of $\sqrt{s}$; $a_s=\alpha_s^{(5)}(m_H)/\pi$.}
    \end{center}
\end{figure}

For illustration we show in Fig.~\ref{fig::gg}
the partonic cross section for the NNLO $n_h^3$ contribution $gg$ channel
as a function of $\sqrt{s}$~\cite{Davies:2019xzc}, where
expansion terms up to ${\cal O}(\delta^{35})$ are included.
This expansion depth is sufficient since we only plot results
up to $\sqrt{s}=500$~GeV.
We observe a reasonable convergence below the top threshold. Above
$\sqrt{s}\approx 350$~GeV the curves diverge which is expected
since the assumed hierarchy $m_t^2\ll s$ does not hold in this region.
Although the region of convergence is small the power corrections provide
important input for the construction of an approximate NNLO result.
For example, in Ref.~\cite{Grober:2017uho} NLO virtual corrections have been
considered and the inclusion of higher order
$1/m_t$ terms significantly stabilizes the Pad\'e results obtained
by combining information from large-$m_t$ and threshold 
expansions.

\section{\label{sec::ae}Asymptotic expansion and building blocks}

We use this section to describe an important cross check of our
calculation, namely the asymptotic expansion performed with {\tt
  exp}~\cite{Harlander:1997zb,Seidensticker:1999bb,q2eexp} as compared
to the ``building-block approach''.

Most of the ingredients needed for the ``building-block approach'' (introduced
in Section~\ref{sec::intro}) can be obtained in a straightforward
way. For example, the four-point function involving two gluons and
two Higgs bosons can be computed by generating the corresponding
four-point amplitude at one and two loops. Since the building block
only contains the hard contribution we can Taylor-expand in the three
independent external momenta and obtain a power series in $1/m_t^2$. In
the numerator we have all possible scalar products, which can be formed
by the external momenta. Note that we have to compute the building
blocks for off-shell gluons and Higgs bosons which makes the results
quite lengthy. For the $ggHH$ building block the colour factor is given
by $\delta^{ab}$ where $a$ and $b$ are the adjoint indices of the
gluons. It can be computed separately since colour and Lorentz parts
factorize. This statement is true for all building blocks involving up
to three gluons but not for the ones involving four gluons.

\begin{figure}[t]
  \begin{center}
    \begin{tabular}{cc}
      \includegraphics[width=0.45\textwidth]{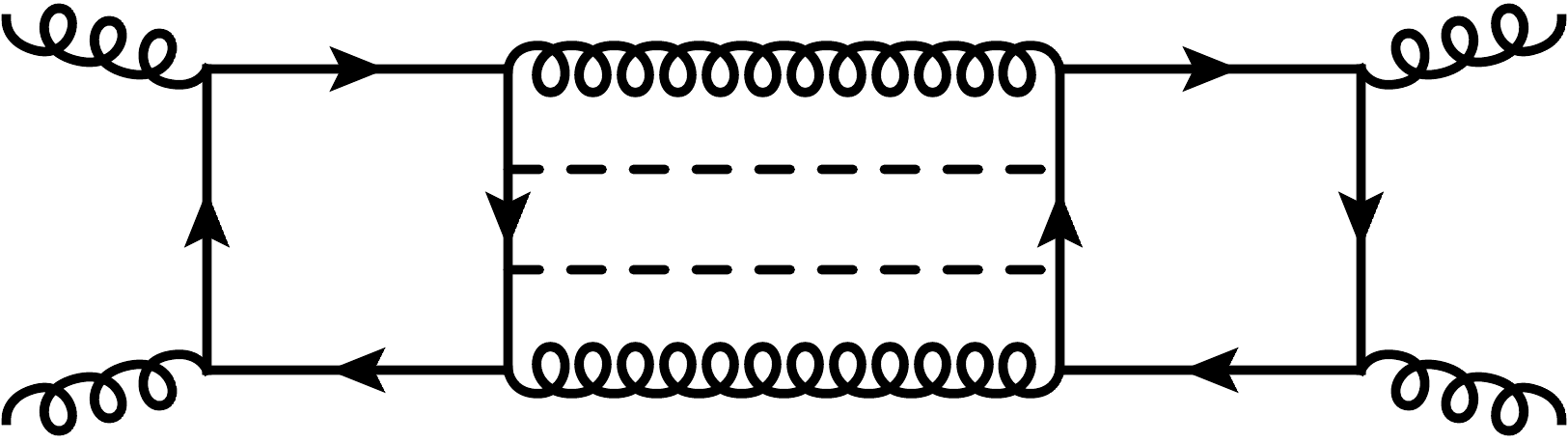}
      &
      \includegraphics[width=0.45\textwidth]{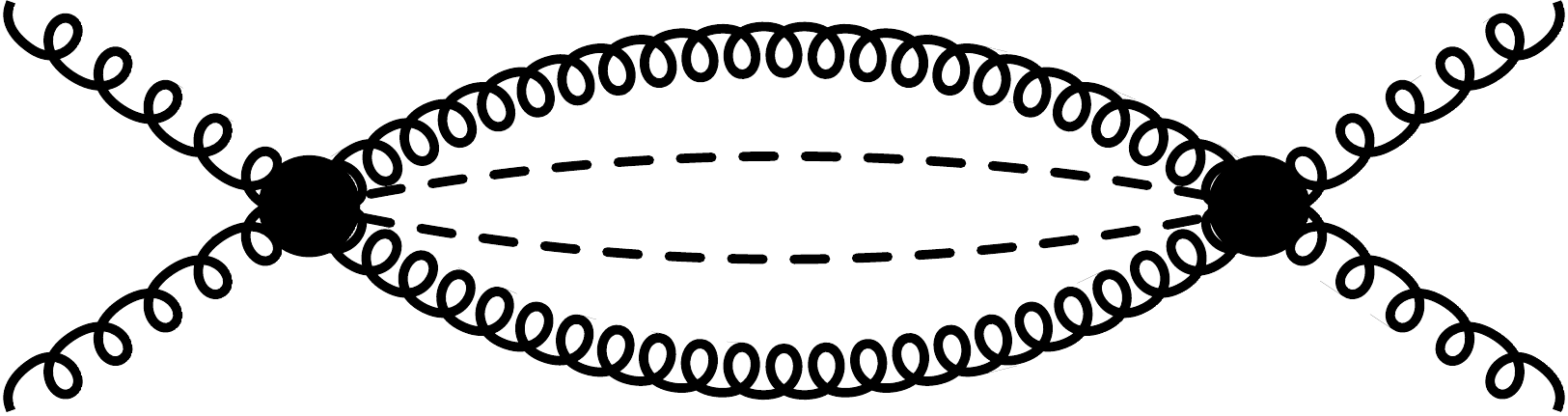}
      \\
      (a) & (b)
    \end{tabular}
  \end{center}
  \caption{\label{fig::4g2h} (a) Sample Feynman diagram contribution to the
    NNLO corrections of $gg\to HH$. (b) Feynman diagram which has to be
    considered in the ``building-block-approach''. The pre-expanded 
    4-gluon-2-Higgs amplitudes have to be inserted at the positions
    of the blobs.
    }
\end{figure}

In the following we consider the class of diagrams which can be described by
the four-gluon-two-Higgs building block. There are 3600 such diagrams\footnote{
Note that the original {\tt qgraf} output contains $1.6\times 10^6$ diagrams.};
a representative five-loop diagram is show in Fig.~\ref{fig::4g2h}(a).
The use of {\tt exp} for this class of diagrams is straightforward; only
one-loop tadpole integrals and a three-loop integral family which corresponds
to Fig.~\ref{fig::4g2h}(b) are needed.
The computation of this small set of diagrams is relatively time consuming,
requiring a wall-time of 70 hours using {\tt TFORM}~\cite{Ruijl:2017dtg} jobs
with four cores.

In the  ``building-block-approach'' approach only one
diagram has to be considered which is shown in Fig.~\ref{fig::4g2h}(b).
Since the Lorentz and colour parts do not
factorize for the amplitudes involving four (off-shell) gluons 
we proceed as follows.

First, we compute the one-loop four-gluon-two-Higgs amplitude 
by simply expanding in the external momenta and arrive
at the result
\begin{align}
  B_{4g2H} = \sum_i K^{\mu\nu\rho\sigma}_i\left(m_t,\{p_j\}\right)C^{abcd}_i~,
\end{align}
where the $K_i$ contain all kinematical information and the $C_i$ are the
colour structures. Note that this quantity has four open Lorentz and four open
colour indices. Next, we introduce the symbols $\delta_i$ with the properties
\begin{align}
  \delta_i\otimes\delta_j = \begin{cases}1 & i = j\,, \\0 & i \neq j\,. \end{cases}
                                                         \label{eq:deltacases}
\end{align}
This allows us to re-write $B_{4g}$ in the form
\begin{align}
B_{4g2H} = \left(\sum_i
  K^{\mu\nu\rho\sigma}_i\left(m_t,\{p_j\}\right)\delta_i\right)
  \otimes
  \left(\sum_j \delta_j C^{abcd}_j\right)\,.
  \label{eq::B4g2H}
\end{align}
In this way we can separate the Lorentz and colour part in the calculation of
the diagrams, which involve the building block $B_{4g2H}$ (see, e.g.,
Fig.~\ref{fig::4g2h}(b)). Both of them contain the quantities $\delta_i$ and
Eq.~(\ref{eq:deltacases}) has to be used when Lorentz and colour part are
multiplied.  Note, that in case we have two building-block insertions with
four gluons in one diagram, we introduce two different sets of $\delta_i$
which commute with each other.

For the diagram shown in Fig.~\ref{fig::4g2h}(b), we obtain for the first two
expansion terms in $1/m_t$
\begin{align}
\mathrm{D_{\mbox{\tiny Fig.~\ref{fig::4g2h}(b)}}} \,\,\sim \,\,& \left(\frac{\alpha_s}{\pi}\right)^4
                          \left(\frac{\mu^2}{m_t^2}\right)^{2\epsilon}
                          \frac{C_A^2T_F^2}{N_A}
                          \frac{
                          \left(2\epsilon+1\right)\left(2\epsilon-3\right) }{ 2\left(1-\epsilon\right)^2 }\nonumber\\
                        &\Bigg\{\left[32\left(2\epsilon+1\right)\left(\epsilon-2\right) +
                          \frac{8m_h^2}{15m_t^2}\left(\epsilon+1\right)\left(10\epsilon^2-31\epsilon-27\right)\right]K_1\nonumber\\
                        &+
                          \frac{8m_h^2}{5m_t^2}\left(\epsilon+1\right)\left(10\epsilon^2-7\epsilon-19\right)K_2\Bigg\}\,,
\end{align}
where $C_A=3$, $T_F=1/2$ and $N_A=8$ are colour factors and $\mu$ is the
renormalization scale. $K_1$ and $K_2$ are master integrals where 
$K_1$ corresponds to the four-particle phase-space (cf. Fig.~\ref{fig::4g2h}(b))
and $K_2$ has an additional numerator of the form
$(m_h^2 - (p_3 + p_4)^2)$, where $p_3$ and $p_4$ are the momenta of the Higgs
bosons.

We use the explicit application of asymptotic expansion to five-loop Feynman
diagrams to cross-check the ``building-block-approach''
at leading order in $1/m_t$. For higher order expansion terms
it becomes quickly very inefficient which is the reason that
we switch to building blocks.


\section*{Acknowledgements}

M.S. would like to thank the organizers of RadCor 2019 for the nice
conference and the pleasant atmosphere.  F.H. acknowledges the
support of the DFG-funded Doctoral School KSETA.  We thank Roman Lee
for the possibility to use the program {\tt LIBRA} and for his
support.  This research was supported by the Deutsche
Forschungsgemeinschaft (DFG, German Research Foundation) under grant
396021762 --- TRR 257 ``Particle Physics Phenomenology after the Higgs
Discovery''.



\end{document}